\documentclass[reprint,amsmath,amssymb,aip,apl]{revtex4-1}

\usepackage{graphicx}
\usepackage{epstopdf}
\usepackage{dcolumn}
\usepackage{bm}
\usepackage{upgreek}

\begin{document}

\preprint{APS/123-QED}

\title{Mode-selected heat flow through a one-dimensional waveguide network}

\author{Christian Riha}
\email[E-mail: ]{riha@physik.hu-berlin.de}

\author{Philipp Miechowski}
\author{Sven S. Buchholz}
\author{Olivio Chiatti}%
\affiliation{%
Novel Materials Group, Humboldt-Universit\"at zu Berlin, 12489 Berlin, Germany
}%

\author{Andreas D. Wieck}
\affiliation{
Angewandte Festk\"orperphysik, Ruhr-Universit\"at Bochum, 44780 Bochum, Germany
}%

\author{Dirk Reuter}
\affiliation{
Angewandte Festk\"orperphysik, Ruhr-Universit\"at Bochum, 44780 Bochum, Germany
}%
\affiliation{%
Optoelektronische Materialien und Bauelemente, Universit\"at Paderborn, 33098 Paderborn, Germany
}%

\author{Saskia F. Fischer}%
\affiliation{%
Novel Materials Group, Humboldt-Universit\"at zu Berlin, 12489 Berlin, Germany
}%

\date{\today}

\begin{abstract}
    Cross-correlated measurements of thermal noise are performed to determine the electron temperature in nanopatterned channels of a GaAs/AlGaAs heterostructure at 4.2~K.
    Two-dimensional (2D) electron reservoirs are connected via an extended one-dimensional (1D) electron waveguide network.
    Hot electrons are produced using a current $I_{\text{h}}$ in a source 2D reservoir, are transmitted through the ballistic 1D waveguide and relax in a drain 2D reservoir.
    We find that the electron temperature increase ${\Delta}T_{\text{e}}$ in the drain is proportional to the square of the heating current $I_{\text{h}}$, as expected from Joule's law.
    No temperature increase is observed in the drain when the 1D waveguide does not transmit electrons.
    Therefore, we conclude that electron-phonon interaction is negligible for heat transport between 2D reservoirs at temperatures below 4.2~K. Furthermore, mode control of the 1D electron waveguide by application of a top-gate voltage reveals that ${\Delta}T_{\text{e}}$ is not proportional to the number of populated subbands $N$, as previously observed in single 1D conductors.
    This can be explained with the splitting of the heat flow in the 1D waveguide network.
\end{abstract}

\maketitle


The transport properties of a one-dimensional (1D) waveguide are dominated by the wave-like character of electrons.
The nanoscale confinement potential is typically created by applying advanced lithographic methods to high mobility two-dimensional electron gases (2DEGs).
In ballistic 1D waveguides electric conductance quantization is observed~\cite{Wees-1988-prl, Wharam-1988-jpc} and is shown to scale linearly with the thermal conductance at low temperatures.~\cite{vanHouten-1992-sst, Chiatti-2006-prl, Jezouin-2013-s}
This indicates the validity of the Wiedemann-Franz relation in the ballistic 1D regime, when electron-phonon and electron-electron interactions can be neglected.~\cite{Sivan-1986-prb, Butcher-1990-jpcm}
In previous works by van Houten~\emph{et al.}~\cite{vanHouten-1992-sst} and Chiatti~\emph{et al.}~\cite{Chiatti-2006-prl} comparable heating measurements between two AlGaAs/GaAs 2DEGs were performed.
The two 2DEGs were connected via a single quantum point contact (QPC) and the increase in electron temperature ${\Delta}T_{\text{e}}$ of the indirectly heated 2D reservoir was measured by a second QPC.
${\Delta}T_{\text{e}}$ was found to be proportional to the number of populated subbands $N$ of the QPC.
The question that arises is how the mode-dependent heat transfer evolves in networks of extended 1D waveguides, where phase-coherent effects have been investigated in- and out-of-equilibrium.~\cite{Buchholz-2012-prb, Chiatti-2014-pssb}

\begin{figure}[b!]
    \includegraphics[width=5cm]{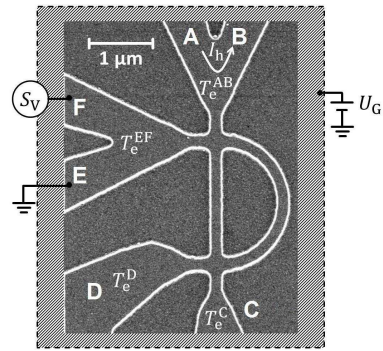}
    \caption{\label{fig:device}
    Scanning electron micrograph of an identically processed sample.
    1D waveguides of about 170~nm lithographic width form an asymmetric ring and are connected to narrow 2D electron reservoirs, labeled A to F.
    The whole structure is covered by a global top-gate.
    $T_{\text{e}}^{\text{AB}}$, $T_{\text{e}}^{\text{EF}}$, $T_{\text{e}}^{\text{C}}$, and $T_{\text{e}}^{\text{D}}$ indicate the electron temperature of the 2D reservoirs and $I_{\text{h}}$ indicates the path of the heating current.
    $S_{\text{V}}$ and $U_{\text{G}}$ indicate the thermal noise and the gate-voltage, respectively.
    }
\end{figure}

Here, we perform cross-correlated electronic noise measurements to determine the charge carrier temperature in an extended 1D waveguide network made from an AlGaAs/GaAs heterostructure.
We investigate the heat transport through an asymmetric quantum ring,~\cite{Chiatti-2014-pssb} which is a network of 1D electron waveguides with 2D contacts as depicted in Fig.~\ref{fig:device}.
A global top-gate enables the control of the conductivity of the 2D reservoirs and the 1D electron waveguides.
One electron reservoir is heated above the lattice temperature via the current heating technique.~\cite{Molenkamp-1990-apl}
The increase in electron temperature ${\Delta}T_{\text{e}}$ of the other electron reservoir is extracted by the means of Johnson-Nyquist noise thermometry.~\cite{Nyquist-1928-pr}
It is a primary thermometry method and is applicable to a wide temperature range.
The temperature of the charge carriers is extracted from thermal noise in resistors independently from the lattice temperature.
Noise thermometry can be applied to bulk material as well as to metal films and wires, due to their diffusive character,~\cite{Roukes-1985-prl, Henny-1997-apl} and to semiconductors hosting high mobility 3D,~\cite{Eckhause-2003-apl} 2D,~\cite{Kurdak-1995-apl, Buchholz-2012-prb} and (quasi-) 1D electronic systems.~\cite{Kurdak-1995-apl}

\begin{figure}[t!]
    \includegraphics[width=7cm]{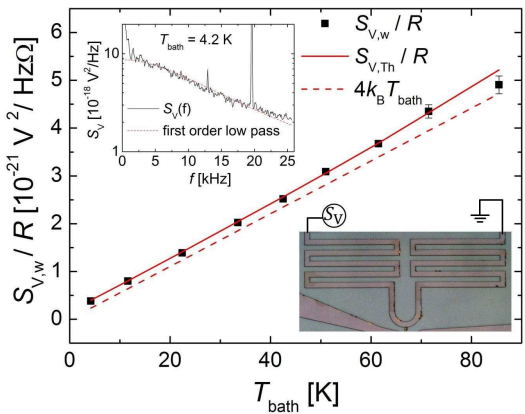}
    \caption{\label{fig:ref-meas}
    Measurements of the (reduced) thermal noise $S_{\text{V,w}}/R$ at different bath temperatures $T_{\text{bath}}$ of a 2DEG from the same AlGaAs/GaAs heterostructure as the sample depicted in Fig.~\ref{fig:device}.
    The (black) squares are the values of $S_{\text{V,w}}/R$ obtained from the noise spectra, the (red) full line is theoretical $S_{\text{V,Th}}/R$ calculated from the two-point resistance $R$ of the sample using Eq.~\ref{eq:total-noise}, and the (red) dashed line is the purely thermal noise, $4k_{\text{B}}T_{\text{bath}}$.
    Lower right inset: optical micrograph of the narrow 2DEG (410~${\upmu}$m length; 2~${\upmu}$m width) where the measurement was performed.
    Upper left inset: noise spectrum at 4.2 K (black line) and the corresponding first-order low-pass fit (red, dashed line) to obtain $S_{\text{V,w}}$.
    }
\end{figure}

Figure~\ref{fig:device} shows a scanning electron microscopy image of a device identical to the one investigated in this work.
It was fabricated from an AlGaAs/GaAs heterostructure with a 2DEG 120~nm below the surface, using electron-beam lithography and wet-chemical etching.
The 2D electron density and mobility at $T = 4.2$~K in the dark are $n = 2.07\times10^{11}$~cm$^{-2}$ and $\mu = 2.43\times10^6$~cm$^2$/Vs, respectively.
These yield a Fermi wavelength of $\lambda_{\text{F}} \approx 55$~nm and a mean free path of $l_{\text{e}} \approx 18~{\upmu}$m.
The 2D electron reservoirs AB and EF are connected with each other via a 1D electron waveguide of 170~nm lithographically defined width.
A narrow 2D channel with 410~$\upmu$m length and 2~${\upmu}$m width (lower inset of Fig.~\ref{fig:ref-meas}), fabricated from the same wafer material, was used to investigate thermal noise in a 2DEG of this heterostructure.

Voltage noise measurements were performed at a bath temperature of $T_{\text{bath}} = 4.2$~K and the noise spectrum $S_{\text{V}}(f)$ was recorded with an \emph{Stanford Research Systems} model SR785 signal analyzer.
At 4.2 K the 2D electron reservoirs yield a thermal noise of the order of  10$^{-18}$~V$^2$/Hz, below the resolution of the signal analyzer.
Two \emph{Signal Recovery} model 5184 low-noise voltage preamplifiers (gain: 10$^3$) were used to increase the thermal noise signal.
Cross-correlated measurements were applied to reduce noise contributions from the preamplifiers.~\cite{Sampietro-1999-rsi}
The noise spectra were taken in a frequency range of $f \approx 1 - 26$~kHz, where 1/f noise is negligible and each noise spectrum $S_{\text{V}}(f)$ is the average of 500 cross-correlated spectra.
In order to take into account parasitic capacitances, each spectrum was fitted with a first-order low-pass filter:
\begin{equation}
\label{eq:low-pass}
    S_{\text{V}}(f) = \frac{S_{\text{V,w}}}{1+(2\pi R C_{\text{par}})^2},
\end{equation}
from which we determine $S_{\text{V,w}}$, the frequency-independent, i.~e. 'white', part of the signal; $R$ is the two-point resistance of the sample, $C_{\text{par}}$ the parasitic capacitance and $k_{\text{B}}$ the Boltzmann constant.
The theoretical value $S_{\text{V,Th}}$ of the measured signal $S_{\text{V,w}}$ is the sum of the thermal noise from the sample, the current-noise of the preamplifiers due to their finite input impedance $R_{\text{amp}}$, and the thermal noise from the leads $S_{\text{V,l}}$:
\begin{equation}
\label{eq:total-noise}
    S_{\text{V,Th}} = 4k_{\text{B}}T_{\text{e}}R + 2 \times 4k_{\text{B}}T_{\text{amp}} \times R^2/R_{\text{amp}} + S_{\text{V,l}}.
\end{equation}
Here $T_{\text{e}}$ denotes the electron temperature, $R_{\text{amp}} = 5$~M$\Omega$, $T_{\text{amp}} \approx 300$~K is the amplifier temperature, and $S_{\text{V,l}} \approx 3\times10^{-19}$~V$^2$/Hz.
Two-point resistance measurements were made by standard lock-in technique using the \emph{Stanford Research Systems} model SR830 lock-in amplifier (frequency $f = 433$~Hz, excitation voltage $U_{\text{ac}} = 40~{\upmu}$V$_{\text{rms}}$), in order to determine $R$.

Figure~\ref{fig:ref-meas} shows the thermal noise of the narrow 2D channel.
The two-point resistance $R$ and the thermal noise $S_{\text{V,w}}$ of the channel were measured in the temperature range of $T_{\text{bath}} = 4.2 - 85$~K; no heating current was applied.
In the range $T_{\text{e}} = 4.2 - 75$~K there is an excellent agreement between the measured $S_{\text{V,w}}$ and the $S_{\text{V,Th}}$ calculated using Eq.~\ref{eq:total-noise}.

\begin{figure}[t!]
    \includegraphics[width=7cm]{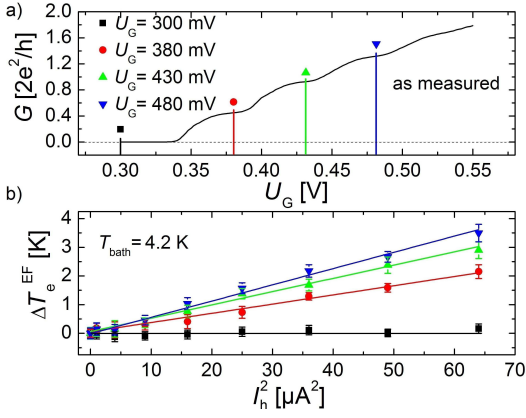}
    \caption{\label{fig:deltaT-vs-Ih}
    Results of the thermal measurements at $T_{\text{bath}} = 4.2$~K.
    (a) Quantized conductance $G$ of the 1D waveguide connecting 2D reservoirs AB and EF, as measured.
    The plateaus appear at the gate-voltages $U_{\text{G}} =$ 380~mV (red circles), 430~mV (green upward triangles) and 480~mV (blue downward triangles), respectively; at $U_{\text{G}} =$ 300~mV (black squares) the 1D waveguide is not conducting.
    (b) Increase in electron temperature ${\Delta}T_{\text{e}}^{\text{EF}}$, calculated using Eq.~\ref{eq:deltaT}, as a function of heating power $P \propto I_\text{h}^2$ at the gate-voltages marked in (a).
    }
\end{figure}

Figure~\ref{fig:deltaT-vs-Ih} shows the results of the measurements using the setup depicted in Fig.~\ref{fig:device}.
The noise spectrum in the 2D electron reservoir EF was measured while applying a heating current $I_{\text{h}}$ to the reservoir AB.
$I_{\text{h}}$ was applied using a battery-driven voltage source with low-pass-filters (each formed by a 1~M$\Omega$ resistor and a 1~${\upmu}$F capacitor) in a ``push-pull'' configuration.
The top-gate was connected to a similar battery-driven voltage source, also with a low-pass-filter (a 100~k$\Omega$ resistor and a 1~${\upmu}$F capacitor).
The ``low'' potential of this battery served as ground.
The electric conductance of the 1D waveguide AB-EF was measured at $T_{\text{bath}} = 4.2$~K in the range $U_{\text{G}} = 0.3 - 0.5$~V.

The number of populated subbands $N$ is given by $N$~= 0, 1, 2, and 3 at $U_{\text{G}}$~= 300~mV, 380~mV, 430~mV, and 480~mV, respectively, as depicted in Fig.~\ref{fig:deltaT-vs-Ih}a; $N = 0$ means that the 1D waveguide AB-EF is not conducting, i.~e. there are no transmitting modes.
For these different occupation numbers, the 2D electron reservoir AB was heated with currents in the range $I_{\text{h}} = 0 - 8~{\upmu}$A in steps of 1~${\upmu}$A. For each $I_{\text{h}}$ the increase in electron temperature ${\Delta}T_{\text{e}}^{\text{EF}}$ in the 2D reservoir EF was determined from the noise spectrum using the following relation:
\begin{equation}
\label{eq:deltaT}
    {\Delta}T_{\text{e}}^{\text{EF}} = \frac{S_{\text{V,w}}(I_{\text{h}}) - S_{\text{V,w}}(I_{\text{h}} = 0)}{4k_{\text{B}}R},
\end{equation}
where $R$ is the two-point resistance of 2D reservoir EF.
30 minutes after each noise measurement at $I_{\text{h}} = 8~{\upmu}$A, the thermal noise was measured again at $I_{\text{h}} = 0~{\upmu}$A to ensure that reservoir EF had cooled down to $T_{\text{bath}}$.
The results of these thermal measurements are presented in Fig.~\ref{fig:deltaT-vs-Ih}b.

Using Eq.~\ref{eq:total-noise} to determine $T_{\text{e}}^{\text{EF}}$ at $I_{\text{h}} = 0~{\upmu}$A yields $T_{\text{e}}^{\text{EF}} \approx 7 - 8$~K, depending on $U_{\text{G}}$; this is higher than $T_{\text{bath}} = 4.2$~K.
However, this difference does not depend on $I_{\text{h}}$ when $U_{\text{G}}$ is constant, as indicated by the ${\Delta}T_{\text{e}}^{\text{EF}} \propto I_{\text{h}}^2$ dependence observed in Fig.~\ref{fig:deltaT-vs-Ih}b.
The increased noise acts as a constant offset for constant gate-voltage and is attributed to capacitively-induced potential fluctuations.

Figure~\ref{fig:deltaT-vs-Ih}b shows that ${\Delta}T_{\text{e}}^{\text{EF}}$ increases with $I_{\text{h}} > 0$ only if electrons transmit through 1D waveguide AB-EF, i.~e. when $N > 1$.
${\Delta}T_{\text{e}}^{\text{EF}} \approx (0.0 \pm 0.2)$~K for any $I_{\text{h}}$ when $N = 0$, which indicates that electron-phonon interaction at $T_{\text{bath}} = 4.2$~K is not strong enough for a direct heat-exchange between the two 2D reservoirs.
Thus, the 2D reservoirs AB and EF are thermally connected only by the 1D electron waveguide.

For $I_{\text{h}} =$~const. the data do not follow the dependence ${\Delta}T_{\text{e}}^{\text{EF}} \propto N$ as previously observed for simple QPCs connecting 2D electron reservoirs.~\cite{vanHouten-1992-sst, Chiatti-2006-prl, Jezouin-2013-s}
To understand this difference, we consider the heat transport in the 1D waveguide network.
In the steady state
\begin{equation}
\label{eq:heat-balance}
        \begin{split}
            \dot{Q}_{\text{AB}} &= \dot{Q}_{\text{EF}}+\dot{Q}_{\text{C}}+\dot{Q}_{\text{D}}+\dot{Q}_{\text{e-ph}}\\
                                &\approx \dot{Q}_{\text{EF}}+\dot{Q}_{\text{C}}+\dot{Q}_{\text{D}},
        \end{split}
\end{equation}
where $\dot{Q}_{\text{AB}}$ denotes the heat flow from reservoir AB into the 1D waveguide network, $\dot{Q}_{\text{EF}}$, $\dot{Q}_{\text{C}}$, and $\dot{Q}_{\text{D}}$ the heat flows into the reservoirs EF, C, and D, respectively, and $\dot{Q}_{\text{e-ph}}$ the heat flow to the lattice due to electron-phonon interaction.
We assume $\dot{Q}_{\text{e-ph}} \approx 0$, because the measurement of ${\Delta}T_{\text{e}}^{\text{EF}}$ for $N = 0$ in Fig.~\ref{fig:deltaT-vs-Ih}b shows that the electron-phonon interaction can be neglected.

\begin{figure}[t!]
    \includegraphics[width=7cm]{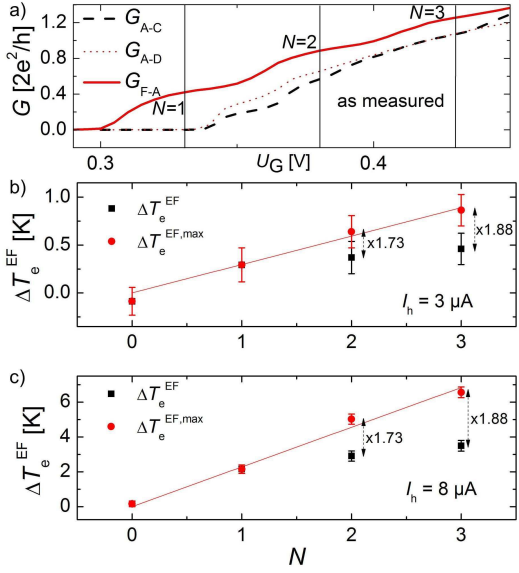}
    \caption{\label{fig:deltaT-max-vs-N}
    Comparison of electric conductances and between the measured ${\Delta}T_{\text{e}}^{\text{EF}}$ and ${\Delta}T_{\text{e}}^{\text{EF,max}}$ calculated with Eq.~\ref{eq:deltaT-max-2}.
    (a) Measured conductance $G$ between contacts A and F (red, full line), A and D (red, dotted line), and A and C (black, dashed line), corresponding to different 1D waveguides.
    The curves are shifted to the left compared to Fig.~\ref{fig:deltaT-vs-Ih}a, because they were measured in a different cooldown; however, the relative position of the curves is the same.
    (b) and (c) Increase in electron temperature ${\Delta}T_{\text{e}}^{\text{EF}}$ (black squares) and ${\Delta}T_{\text{e}}^{\text{EF,max}}$ (red circles) as a function of the number of populated subbands, for $I_{\text{h}} = 3~\upmu$A and $I_{\text{h}} = 8~\upmu$A, respectively.
    }
\end{figure}

The heat flow through the different paths in the 1D waveguide network can be expressed as follows:
\begin{equation}
\label{eq:heat-flows}
    \begin{split}
        \dot{Q}_{\text{EF}} &= \kappa_{\text{AB-EF}}(T_{\text{e}}^{\text{AB}}-T_{\text{e}}^{\text{EF}})\\
                            &= \kappa_{\text{AB-EF}} \left[(T_{\text{L}} + {\Delta}T_{\text{e}}^{\text{AB}}) - (T_{\text{L}} + {\Delta}T_{\text{e}}^{\text{EF}})\right]\\
                            &= \kappa_{\text{AB-EF}} \left[{\Delta}T_{\text{e}}^{\text{AB}} - {\Delta}T_{\text{e}}^{\text{EF}}\right]\\
        \dot{Q}_{\text{C}}  &= \kappa_{\text{AB-C}}\left[{\Delta}T_{\text{e}}^{\text{AB}} - {\Delta}T_{\text{e}}^{\text{C}}\right]\\
        \dot{Q}_{\text{D}}  &= \kappa_{\text{AB-D}}\left[{\Delta}T_{\text{e}}^{\text{AB}} - {\Delta}T_{\text{e}}^{\text{D}}\right],
    \end{split}
\end{equation}
where ${\Delta}T_{\text{e}}^{\text{X}}$ describes the increase of the electron temperature over the lattice temperature $T_{\text{L}}$ in reservoir X and $\kappa_{\text{X-Y}}$ is the thermal conductance of the 1D waveguide from reservoir X to reservoir Y.
The reservoirs C and D are on the same side of the ring structure and are separated from reservoirs AB and EF by the same 1D waveguide.
Conductance measurements (see Fig.~\ref{fig:deltaT-max-vs-N}(a) show that $G_{\text{AB-C}} \approx G_{\text{AB-D}}$, so we can assume that ${\Delta}T_{\text{e}}^{\text{C}} \approx {\Delta}T_{\text{e}}^{\text{D}} \equiv {\Delta}T_{\text{e}}^{\text{CD}}$ and write $\kappa_{\text{AB-CD}} = \kappa_{\text{AB-C}} + \kappa_{\text{AB-D}}$.
Defining $\dot{Q}_{\text{CD}} \equiv \dot{Q}_{\text{C}} + \dot{Q}_{\text{D}}$, the ratio $\dot{Q}_{\text{CD}}/\dot{Q}_{\text{EF}}$ is then
\begin{equation}
\label{eq:heat-ratio-1}
    \frac{\dot{Q}_{\text{CD}}}{\dot{Q}_{\text{EF}}}=\frac{\kappa_\text{AB-CD}(\Delta T_\text{e}^{\text{AB}}-\Delta T_\text{e}^{\text{CD}})}{\kappa_\text{AB-EF}(\Delta T_\text{e}^{\text{AB}}-\Delta T_\text{e}^{\text{EF}})}.
\end{equation}
Applying the Wiedemann-Franz relation yields
\begin{equation}
\label{eq:therm-conds}
    \begin{split}
        \kappa_{\text{AB-EF}}   &= LG_{\text{AB-EF}}(T_{\text{e}}^{\text{AB}} + T_{\text{e}}^{\text{EF}})/2 =\\
                                &= LG_{\text{AB-EF}} \left[(T_{\text{L}} + {\Delta}T_{\text{e}}^{\text{AB}}) + (T_{\text{L}} + {\Delta}T_{\text{e}}^{\text{EF}})\right]/2\\
                                &= LG_{\text{AB-EF}} \left[2 T_{\text{L}} + {\Delta}T_{\text{e}}^{\text{AB}} + {\Delta}T_{\text{e}}^{\text{EF}}\right]/2\\
        \kappa_{\text{AB-CD}}   &= LG_{\text{AB-CD}} \left[2 T_{\text{L}} + {\Delta}T_{\text{e}}^{\text{AB}} + {\Delta}T_{\text{e}}^{\text{CD}}\right]/2,
    \end{split}
\end{equation}
where $L$ is the Lorenz number and $G_{\text{X-Y}}$ the electric conductances between reservoirs X and Y.
Combining Eqs.~\ref{eq:heat-ratio-1} and~\ref{eq:therm-conds} leads to
\begin{equation}
\label{eq:heat-ratio-2}
    \begin{split}
        \frac{\dot{Q}_{\text{CD}}}{\dot{Q}_{\text{EF}}} = \frac{G_{\text{AB-CD}}}{G_{\text{AB-EF}}} &\times \frac{(2 T_{\text{L}} + {\Delta}T_{\text{e}}^{\text{AB}} + {\Delta}T_{\text{e}}^{\text{CD}})}{(2 T_{\text{L}} + {\Delta}T_{\text{e}}^{\text{AB}} + {\Delta}T_{\text{e}}^{\text{EF}})}\\
                                                        &\times \frac{({\Delta}T_{\text{e}}^{\text{AB}} - {\Delta}T_{\text{e}}^{\text{CD}})}{({\Delta}T_{\text{e}}^{\text{AB}} - {\Delta}T_{\text{e}}^{\text{EF}})}.
    \end{split}
\end{equation}
In Fig.~\ref{fig:deltaT-vs-Ih}b it can be seen that for a small heating current, $I_{\text{h}} \lesssim 3~{\upmu}$A, the rise in the electron temperature of reservoir EF is much smaller than the lattice temperature, ${\Delta}T_{\text{e}}^{\text{EF}} \ll T_{\text{L}}$.
So, with ${\Delta}T_{\text{e}}^{\text{EF}}$, ${\Delta}T_{\text{e}}^{\text{CD}} \ll T_{\text{L}}$, ${\Delta}T_{\text{e}}^{\text{AB}}$ Eq.~\ref{eq:heat-ratio-2} becomes
\begin{equation}
\label{eq:heat-ratio-3}
    \frac{\dot{Q}_{\text{CD}}}{\dot{Q}_{\text{EF}}} \approx \frac{G_{\text{AB-CD}}}{G_{\text{AB-EF}}} \equiv x,
\end{equation}
where $x$ is the ratio of the electric two-point conductances between different reservoirs.

The maximum heat flow to the reservoir EF can be estimated by setting $\dot{Q}_{\text{CD}} = 0$
\begin{equation}
\label{eq:heat-flow-max-1}
    \dot{Q}_{\text{AB}}=\dot{Q}_{\text{EF}}^{\text{max}} = \kappa_{\text{AB-EF}}^{\text{max}}\left[{\Delta}T_{\text{e}}^{\text{AB}} - {\Delta}T_{\text{e}}^{\text{EF,max}}\right],
\end{equation}
with
\begin{equation}
\label{eq:therm-cond-max-1}
    \kappa_{\text{AB-EF}}^{\text{max}} = L G_{\text{AB-EF}}^{\text{max}}\left[ 2 T_{\text{L}} + {\Delta}T_{\text{e}}^{\text{AB}} + {\Delta}T_{\text{e}}^{\text{EF,max}}\right]/2.
\end{equation}
Here, $G_{\text{AB-EF}}^{\text{max}} = G_{\text{AB-EF}}$, because $U_{\text{G}}$ is constant.
Therefore, the heat balance, Eq.~\ref{eq:heat-balance}, can be rewritten using Eqs.~\ref{eq:heat-ratio-3} and~\ref{eq:heat-flow-max-1} as follows:
\begin{equation}
\label{eq:heat-flow-max-2}
    \dot{Q}_{\text{EF}}^{\text{max}} \approx \dot{Q}_{\text{EF}} + \dot{Q}_{\text{CD}} \approx \dot{Q}_{\text{EF}}(1+x).
\end{equation}
Using Eqs.~\ref{eq:heat-flows} and~\ref{eq:heat-flow-max-1} yields
\begin{equation}
\label{eq:therm-cond-max-2}
    \kappa_{\text{AB-EF}}^{\text{max}} {\Delta}T_{\text{e}}^{\text{EF,max}} \approx \kappa_{\text{AB-EF}} {\Delta}T_{\text{e}}^{\text{EF}}(1+x);
\end{equation}
$\kappa_{\text{AB-EF}}^{\text{max}}$ and $\kappa_{\text{AB-EF}}$ can be replaced using Eqs.~\ref{eq:therm-conds} and~\ref{eq:therm-cond-max-1}, respectively, and we obtain
\begin{equation}
\label{eq:deltaT-max-1}
    \begin{split}
        \left(2 T_{\text{L}} + {\Delta}T_{\text{e}}^{\text{AB}} + {\Delta}T_{\text{e}}^{\text{EF,max}} \right) \left({\Delta}T_{\text{e}}^{\text{AB}} - {\Delta}T_{\text{e}}^{\text{EF,max}} \right) \approx\\
        \left(2 T_{\text{L}} + {\Delta}T_{\text{e}}^{\text{AB}} + {\Delta}T_{\text{e}}^{\text{EF}} \right) \left({\Delta}T_{\text{e}}^{\text{AB}} - {\Delta}T_{\text{e}}^{\text{EF}} \right)(1+x).
    \end{split}
\end{equation}
The approximation ${\Delta}T_{\text{e}}^{\text{EF}}$, ${\Delta}T_{\text{e}}^{\text{CD}} \ll T_{\text{L}}$, ${\Delta}T_{\text{e}}^{\text{AB}}$ yields finally
\begin{equation}
\label{eq:deltaT-max-2}
    {\Delta}T_{\text{e}}^{\text{EF,max}} \approx {\Delta}T_{\text{e}}^{\text{EF}}(1+x) = {\Delta}T_{\text{e}}^{\text{EF}}\left(1 + \frac{G_{\text{AB-CD}}}{G_{\text{AB-EF}}}\right).
\end{equation}

Figure~\ref{fig:deltaT-max-vs-N} shows a comparison between the measured ${\Delta}T_{\text{e}}^{\text{EF}}$ and ${\Delta}T_{\text{e}}^{\text{EF,max}}$ calculated with Eq.~\ref{eq:deltaT-max-2}.
The electric two-point conductances $G_{\text{F-A}} \approx G_{\text{AB-EF}}$, $G_{\text{A-C}} \approx G_{\text{AB-CD}}$ and $G_{\text{A-D}} \approx {G_\text{AB-CD}}$ are dominated by the 1D waveguides and were measured to determine $x$ for different values of $U_{\text{G}}$, i.~e. for different $N$.
The ratio $x$ was found to be $x = 0$ for $N = 0$, $x \approx 0$ for $N = 1$, $x \approx 0.73$ for $N = 2$, and $x \approx 0.88$ for $N = 3$ (see Fig.~\ref{fig:deltaT-max-vs-N}a).
The comparison for $I_{\text{h}} = 3~\upmu$A, i.~e. for the linear regime, is shown in Fig.~\ref{fig:deltaT-max-vs-N}b.
However, Fig.~\ref{fig:deltaT-max-vs-N}c shows that Eq.~\ref{eq:deltaT-max-2} holds also for the non-linear regime.~\cite{Chiatti-2006-prl}

To conclude, the heat flow splits for $N > 1$ in the quantum wire network and prevents the observation of ${\Delta}T_{\text{e}} \propto N$, valid for a single QPC.~\cite{vanHouten-1992-sst, Chiatti-2006-prl}
The difference of these results can be explained by taking into account the heat flow along different paths and can be estimated from the electric conductance of the different paths.
The estimate is based on the assumption that the Wiedemann-Franz relation holds~\cite{vanHouten-1992-sst, Chiatti-2006-prl, Jezouin-2013-s} and  the ratio of the electric conductances of the 1D waveguides in the network determines the temperature increase ${\Delta}T_{\text{e}}^{\text{EF}} = {\Delta}T_{\text{e}}^{\text{EF}}(N)$ in the reservoir EF.
The top-gate voltage controls which 1D modes carry heat across the structure and therefore allows a selective heating of the 2D reservoirs connected to the quantum wire network.
This has relevance for future applications, such as quantum circuits made of extended electron waveguide networks. 

We gratefully acknowledge financial support by the priority programme ``Nanostructured thermoelectrics''� of the German Science Foundation (DFG) SPP 1386 grant Nr. Fi932/2-2. A.~D.~W. acknowledges gratefully support of Mercur  Pr-2013-0001, BMBF-Q.com-H 16KIS0109, and the DFH/UFA CDFA-05-06. We further thank Dr. R\"udiger Mitdank, Dr. Tobias Kramer and Dr. Christoph Kreisbeck for fruitful scientific discussions.

\bibliography{Rauschen-Paper}

\end{document}